\documentclass[prfluids,twocolumn]{revtex4-2}
\usepackage{amsfonts}
\usepackage{amsmath}
\usepackage{amssymb}
\usepackage{graphicx}
\usepackage{bm}
\usepackage{xcolor}

\renewcommand{\d}{\text{d}}

\begin{document}

\title{Blockage of thermocapillary flows by surface-active impurities}
\author{Thomas Bickel}
\email{thomas.bickel@u-bordeaux.fr}
\affiliation{Univ. Bordeaux \& CNRS, LOMA (UMR 5798),  F-33400 Talence, France}

\begin{abstract}
Thermocapillary convection is particularly effective for the control of thin liquid film topography or for the actuation of microparticles at the liquid-air interface. Experiments with water are challenging, however, as its interface is prone to contamination by surface-active impurities that profoundly alter its hydrodynamic response.
Despite numerous reports highlighting the hindrance of thermocapillary flows, quantitative information on interface contamination is definitely lacking. 
We therefore introduce a general framework in order to account for the presence of a low concentration of insoluble surfactants at the water-air interface.
Focusing on the low-compressibility limit, we identify the inverse Marangoni number as the appropriate small parameter in the modeling. The transport equations can then be linearized without any further assumption regarding the other dimensionless groups. It is shown in particular that the surfactant concentration adapts to the thermocapillary forcing in order to cancel the interfacial viscous stress.
Several experimentally relevant situations are discussed  with emphasis on physical observables. Besides the excellent agreement between theory and experiment, our predictions provide a quantitative assessment of the impurity concentration. 
\end{abstract}

\maketitle

\section{Introduction}
\label{sec_intro}

The physical properties of the liquid-gas interface are primarily controlled by its surface tension. For most simple liquids, the  latter is  a decreasing function of the temperature. When a non-uniform heating is established, the resulting Marangoni stresses induce a thermocapillary flow from warmer regions  (with lower surface tension) to colder regions (with higher surface tension). Thermocapillary convection was for instance exploited for the propulsion of microparticles~\cite{darhuberARFM2005,karbalaeiMuMa2016,chenChemRev2022}. More specifically, the rotational motion of floating microgears was  observed under LED illumination~\cite{maggiNatComm2015}, while photothermal actuation was achieved for spherical microparticles~\cite{girotLangmuir2016}. It was further demonstrated that the oscillatory motion of an assembly of particles  can be controlled using patterned illumination~\cite{haywardPRL2018,haywardPNAS2021,haywardSoftMat2024}.  Other applications  include droplet manipulation~\cite{baroudLabC2007} or shaping  of thin liquid films with a predetermined topography~\cite{rubinLSA2019,ivanovaColInt2022,eshelFlow2022}. The thermocapillary effect also proved to be a valuable tool to characterize the thermophysical properties of complex fluids~\cite{vermaLSA2022}. From a more fundamental standpoint, optically-induced thermal gradients were exploited to control the thickness of thin films  down to the nanoscale, which allowed to probe the contribution of van der Waals forces on interfacial fluctuations~\cite{clavaudPRL2021} or deformations~\cite{mazaPRF2022}.

The ability to control and manipulate the liquid-air interface requires a precise understanding of the response to temperature variations. The purity of the system is therefore of paramount importance. For this reason, cautious experiments are usually done with low surface energy liquids such as silicon oils. In contrast, experiments with water are generally more involved since its interface is prone to contamination by organic impurities. Such surface-active molecules are inevitably present in the surroundings and have the unfortunate tendency to absorb in an uncontrolled manner at the water-air interface. Contrary to what can be presumed, the change in surface tension doesn't really matter in the first place. Indeed, the surface tension remains almost unaffected at low impurity concentration. However, the surface tension \emph{gradients}, which emerge from the compression of the surfactant-laden interface, are known to profoundly alter the hydrodynamic response of the liquid layer~\cite{lohseJFM2023}. The hydrodynamic boundary condition can indeed switch locally from no-stress to no-slip due to the presence of a minute amount of contaminants~\cite{levichbook,manikantanJFM2020}. This explains for instance why the ascending velocity of a gas bubble in water is almost always lower than the theoretical prediction assuming a stress-free condition~\cite{palaparthiJFM2006}. The reason for this retardation is that adsorbed surfactant is swept to the trailing pole of the bubble, where it accumulates and lowers the surface tension relative to the front end. The resulting Marangoni forces oppose the surface flow and immobilize the interface, thus increasing the friction force. This stagnant cap effect is also responsible for, \textit{e.g.}, the unwanted drag increase near superhydrophobic surfaces~\cite{peaudecerfPNAS2017,tomlinsonJFM2023} or the arrest of interfacial divergent flows~\cite{bickelPRF2019,koleskiPRF2021}. 

In the context of thermocapillary convection, it has long been recognized that surface immobilization can be used to unveil the presence of traces of organic impurities~\cite{bezuglyiCSA2004}. Several studies recently tried to elucidate the complex interplay between thermal, viscous and molecular transport~\cite{vinnichenkoIJHMT2018,bickelEPJE2019,koleskiPoF2020,pinanPoF2021,rudenkoJFM2022}, yet an overall understanding is still elusive. In particular, no quantitative information regarding the features of the contaminants has been obtained so far.
The purpose of the present paper is to identify the ``hidden'' parameters pertaining to the contaminants~\cite{manikantanJFM2020}. More specifically, we aim at characterizing the hydrodynamic signature of  surface-active impurities on thermocapillary flows. This program needs to combine  experimentation and theory in a comprehensive study. From this viewpoint, numerical models with too many free parameters are not really helpful. Instead, emphasis is placed on key physical features in order to come up with an approach that is analytically tractable.

The concept of stagnant cap was first adapted to thermocapillary flows by Homsy and collaborators~\cite{homsyJFM1984,carpenterJFM1985}. In this seminal series of papers, the authors focused on the 2D thermocapillary flow resulting from a linear temperature profile along the interface. However, the extension of the immobile region, as theoretically predicted, does not coincide with experimental observations~\cite{shmyrovJFM2019}.  The analysis was further generalized to the 3D axisymmetric configuration, but here again the observed scaling for the large-distance decay of the velocity does not fit with theoretical predictions~\cite{bickelEPJE2019,koleskiPoF2020}. In the present work, we re-examine this question from a new perspective, which requires rethinking commonly made assumptions.
This step is essential because the material parameters relating to the surfactants are largely unknown. Our analysis closely follows the line of reasoning developed by Shardt, Masoud and Stone~\cite{shardtJFM2016}. Basically, the idea is that the surfactant layer is almost incompressible even in very dilute regime~\cite{manikantanJFM2020}.  Focusing on the low-compressibility limit, we identify the inverse Marangoni number as the appropriate small parameter.  The transport equations can then be linearized without any further assumption regarding the other dimensionless parameters  (Reynolds, Péclet, Biot,  \ldots) that are involved in the theoretical description. From then on, the problem becomes analytically tractable, at least in principle. Our general framework can thus be applied the various experimentally relevant situations, some of which been discussed in this work.  

The remaining of the paper is organized as follows. The mathematical problem is first formulated in Sec.~\ref{sec_eq}. The limit of low-compressibility is then introduced in Sec.~\ref{sec_inc}, and the steady-state solutions of the linearized transport equations are discussed in Sec.~\ref{seq_steady}. Three illustrating examples are presented: the 2D flow resulting from a line heat source, and the 3D axisymmetric flows due to heating by a point-like source and a laser beam.  The main outcome of this work are finally discussed in Section~\ref{sec_disc}. In particular, it is shown how the concentration of surface-active impurities can be extracted from experimental data.

\section{Mathematical formulation}
\label{sec_eq}

%%%%%%%%%%%%%%%%%%%%%%%%%%%%%%%%%%%%%%%
\begin{figure}
\centering
\includegraphics[width=\columnwidth]{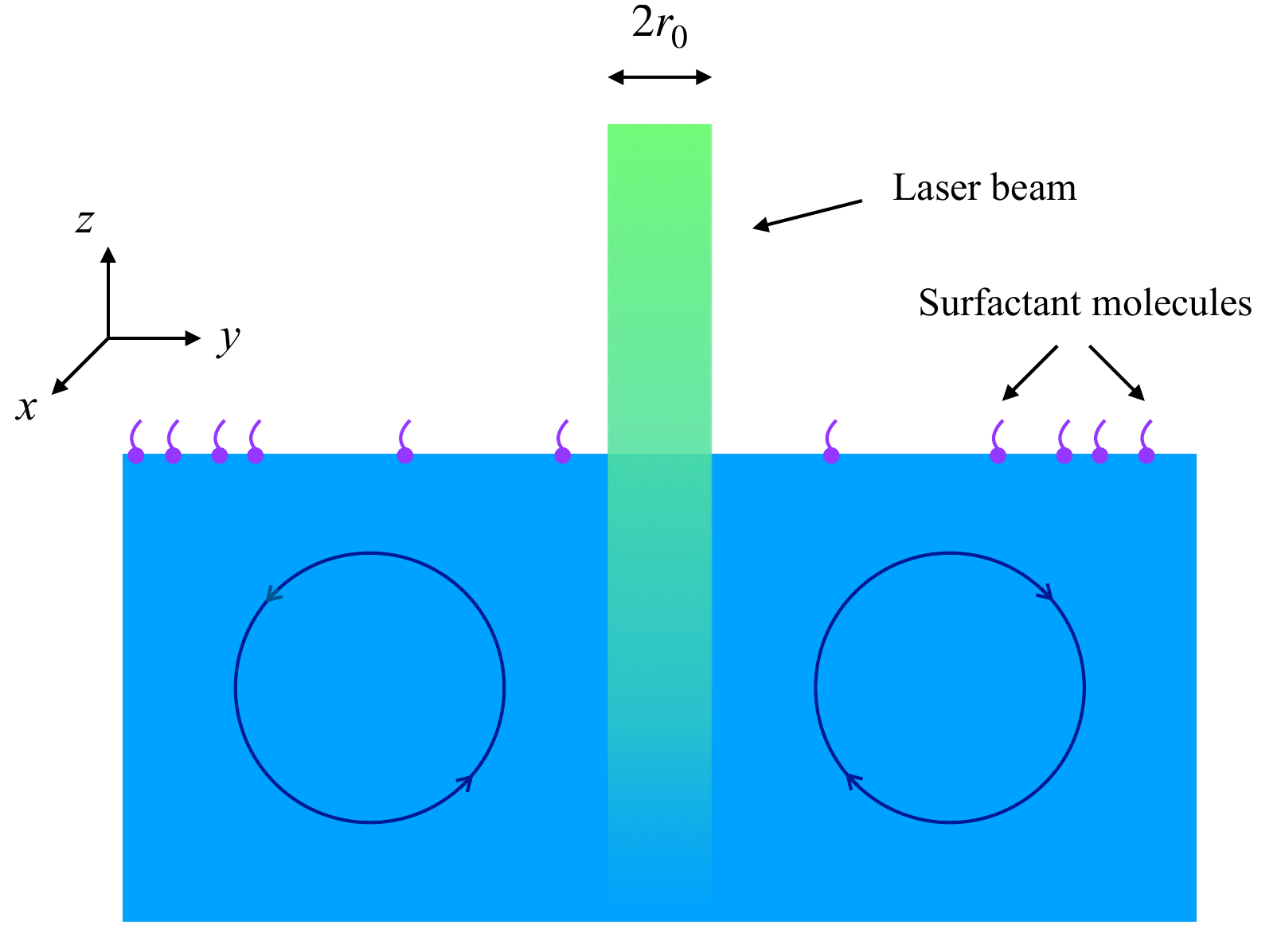}
\caption{An inhomogeneous temperature field drives a thermocapillary flow in the liquid phase. In this schematic, the liquid is heated by an axisymmetric laser beam of waist $r_0$. The flow is directed towards colder regions, with recirculation at distances of the order of the size of the container. When a low amount of surface-active impurities is present, the resulting solutal counterflow strongly modifies the features of the original thermocapillary flow.}
\label{schema}       
\end{figure}
%%%%%%%%%%%%%%%%%%%%%%%%%%%%%%%%%%%%%%%

The situation that we aim to describe is depicted on Fig.~\ref{schema}. We consider an incompressible, Newtonian liquid which is  quiescent under uniform temperature~$T_0$. At equilibrium, the liquid-air interface is planar and coincides with the horizontal $xOy$ plane, the vertical $z$-axis been oriented upward. The system is then driven out of equilibrium by switching on a local heat source. As a  consequence of the temperature gradient, Marangoni stresses build up at the interface and set the liquid into motion by thermocapillary effect. The resulting  interfacial flow is directed away from the heat source towards the colder regions.

The features of the flow are strongly modified when surface-active impurities are adsorbed at the interface.
In this case, the thermally-driven flow sweeps the molecules away from the heat source. The result is a concentration gradient of surfactant that in turn causes a secondary Marangoni flow, of solutal origin this time, which opposes the initial thermal flow. Our aim is to elucidate the competition between the outward thermocapillary flow and the reverse solutocapillary flow. The ultimate goal of this study is to unveil the signatures of these otherwise undetectable impurities by focusing on the properties of the hindered interfacial flow. 

The main difficulty when dealing with non-isothermal flows is that the physical properties of the liquid --- viscosity~$\eta$, mass density~$\rho$, heat capacity~$c$, heat conductivity~$\kappa$, surface tension~$\gamma$ --- are likely to depend on the local temperature~$T$.
Throughout this paper, we adopt  Boussinesq standpoint and assume that all material parameters are constant, except for the surface tension in the Marangoni boundary condition and the buoyancy term in the Navier-Stokes equation. For finite deviation with respect to $T_0$, the liquid mass density can written
%%%%%%%%%%%%%%%%%%%%%%%%%%%%%%%%%%%%%%%%%%%
\begin{equation}
\rho(T)\approx \rho_0\left[ 1-\alpha (T-T_0) \right]  \ ,
\end{equation}
%%%%%%%%%%%%%%%%%%%%%%%%%%%%%%%%%%%%%%%%%%%
with $\alpha=\vert \rho^{-1} \partial \rho / \partial T\vert$  the coefficient of thermal expansion evaluated at $T=T_0$, and $\rho_0=\rho(T_0)$.  Regarding the surface tension, it is for most simple liquids  a decreasing function of both the temperature~$T$ and the surface concentration~$\Gamma$. 
One can assume a linear relationship for small variations around  equilibrium 
%%%%%%%%%%%%%%%%%%%%%%%%%%%%%%%%%%%%%%%%%%%
\begin{equation}
\gamma (T, \Gamma) \approx \gamma_0 - \gamma_{\theta}(T-T_0)  - \gamma_s(\Gamma - \Gamma_0)    \ ,
\label{gamma}
\end{equation}
%%%%%%%%%%%%%%%%%%%%%%%%%%%%%%%%%%%%%%%%%%%
where the Marangoni coefficients $\gamma_{\theta}= \vert \partial \gamma / \partial T\vert $ and $\gamma_{s}=\vert \partial \gamma / \partial \Gamma\vert $ are  positive constants.   Here, $\gamma_0=\gamma (T_0, \Gamma_0)$ is the reference value pertaining to the initially uniform concentration of surfactant~$\Gamma_0$.

We start by specifying the transport equations in the liquid subphase ($z\leq 0$). The conservation of momentum  is expressed by the Navier-Stokes equation 
%%%%%%%%%%%%%%%%%%%%%%%%%%%%%%%%%%%%%%%%%%%
\begin{align}
\rho_0 \left( \frac{\partial \mathbf{v}}{\partial t} + \mathbf{v}\cdot \bm{\nabla} \mathbf{v} \right) =  \eta \nabla^2 \mathbf{v} -  & \bm{\nabla} (P+\rho_0 g z) \nonumber \\ & +\rho_0 g \alpha (T-T_0) \mathbf{e}_z \ ,
\label{momentum}
\end{align}
%%%%%%%%%%%%%%%%%%%%%%%%%%%%%%%%%%%%%%%%%%%
where $\mathbf{v}=(v_x,v_y,v_z)$ is the velocity and  $P$ the pressure.
This equation has to be solved together with the incompressibility condition
%%%%%%%%%%%%%%%%%%%%%%%%%%%%%%%%%%%%%%%%%%%
\begin{equation}
\bm{\nabla} \cdot \mathbf{v}=0  \ .
\label{incomp}
\end{equation}
%%%%%%%%%%%%%%%%%%%%%%%%%%%%%%%%%%%%%%%%%%%
Energy conservation  is then enforced by the heat equation 
%%%%%%%%%%%%%%%%%%%%%%%%%%%%%%%%%%%%%%%%%%%
\begin{equation}
\rho_0 c \left( \frac{\partial T}{\partial t} + \mathbf{v}\cdot \bm{\nabla} T \right) = \kappa \nabla^2 T + q(x,y,z,t)  \ ,
\label{energy}
\end{equation}
%%%%%%%%%%%%%%%%%%%%%%%%%%%%%%%%%%%%%%%%%%%
with $q$ the injected power density. The latter depends on the specifics of the experimental situation.
Three cases are discussed in this paper, with a particular focus on the steady-state regime. First, we  consider the flow induced by a line heat source located at some distance $H$ below the free surface. Experimentally, heating can be achieved using a thin metal wire connected to a DC power supply. If the diameter of the wire is much smaller than all other length scales, the power density  can be written [case 1]
%%%%%%%%%%%%%%%%%%%%%%%%%%%%%%%%%%%%%%%%%%%
\begin{equation}
q (x,z) = Q_l \delta (x)  \delta(z+H)  \ ,
\label{defq1}
\end{equation}
%%%%%%%%%%%%%%%%%%%%%%%%%%%%%%%%%%%%%%%%%%%
with $Q_l$ the injected power per unit length. Defining $\Delta T \doteq Q_l/(2\pi \kappa)$,
one therefore expects a temperature increase of the order of $\Delta T  \approx 8~\text{K}$ for a supply power  $Q_l \approx 30~\text{W}\, \text{m}^{-1}$ and $\kappa \approx 0.6~\text{W} \, \text{m}^{-1} \, \text{K}^{-1}$. This indeed corresponds to the order of magnitude  reported in the literature~\cite{vinnichenkoIJHMT2018,rudenkoJFM2022,shahdhaarPoF2023}. As a second instance, we  focus on the thermocapillary flow due to a point-like heat source  located at the free surface. This is achieved for instance with a small spherical particle that lies in partial wetting condition at the water-air interface~\cite{girotLangmuir2016,koleskiPoF2020}. Under laser illumination, the particle adsorbs the light energy and releases heat in the liquid. Assuming again that the particle radius~$a$ is much smaller than all the other length scales, the power density reads [case 2] 
%%%%%%%%%%%%%%%%%%%%%%%%%%%%%%%%%%%%%%%%%%%
\begin{equation}
q (r,z) = \frac{Q}{2\pi r} \delta (r)  \delta(z)  \ ,
\label{defq3}
\end{equation}
%%%%%%%%%%%%%%%%%%%%%%%%%%%%%%%%%%%%%%%%%%%
with $Q$ the total injected power, and $r=\vert \mathbf{r}\vert =\sqrt{x^2+y^2}$. For  $Q\approx 4~\text{mW}$ and  $a \approx 100~\mu \text{m}$, one thus anticipates a temperature increase  $\Delta T\doteq Q/(2\pi \kappa a) \approx 10~\text{K}$, as is indeed observed experimentally. As a third application, we investigate the photothermal heating by a laser beam that propagates downward along the vertical axis. Assuming a Gaussian beam profile, the power density has the following form [case 3]  
%%%%%%%%%%%%%%%%%%%%%%%%%%%%%%%%%%%%%%%%%%%
\begin{equation}
q (r,z) = \frac{2kQ}{\pi    r_0^3}\,  e^{z/z_0}  e^{-r^2/(2r_0^2)}   \ ,
\label{defq2}
\end{equation}
%%%%%%%%%%%%%%%%%%%%%%%%%%%%%%%%%%%%%%%%%%%
with $Q$ the laser power, $r_0$ the beam waist, and $z_0$ the penetration length. We also set $k=r_0/z_0$. The temperature scale is then defined as $\Delta T \doteq 2Q/(\pi \kappa r_0)$. In the experimental part of Ref.~\cite{pinanPoF2021}, both $z_0$ and $r_0$ are in the submillimeter range and the maximum temperature elevation varies between 2~K to 8~K.

The remaining transport equation concerns  surface-active impurities. The latter are described as insoluble surfactants that are dispersed at the liquid-air interface. 
Since the interface may deform under hydrodynamic stresses, the mass conservation equation is known to couple the local surfactant concentration~$\Gamma$ to  surface deformations. Quite generally, the location of the interface can be described by a function $\Phi$ such that $\Phi(x,y,z,t)=0$ for any point that belongs to the interface. The unit normal vector is then defined as $\mathbf{n} = \bm{\nabla} \Phi / \vert \bm{\nabla} \Phi \vert$. The advection-diffusion equation for surfactant transport along a deforming interface thus reads~\cite{stonePoF1990}
%%%%%%%%%%%%%%%%%%%%%%%%%%%%%%%%%%%%%%%%%%%
\begin{equation}
\frac{\partial \Gamma}{\partial t}  + \bm{\nabla}_{s} \cdot \left(  \Gamma \, \mathbf{v}_s \right)+\Gamma \left( \bm{\nabla}_s \cdot \mathbf{n}\right) (\mathbf{v} \cdot \mathbf{n} )  = D_s \nabla_{s}^2 \Gamma   \ ,
\label{advdiff}
\end{equation}
%%%%%%%%%%%%%%%%%%%%%%%%%%%%%%%%%%%%%%%%%%%
with $D_s$ the surface diffusion coefficient. Here, the $s$ subscript denotes projection onto the deformed interface, \textit{e.g.}, $\bm{\nabla}_{s}= \left( \bm{I} - \mathbf{n} \mathbf{n} \right) \cdot \bm{\nabla} $ for the surface gradient operator, and $ \mathbf{v}_s = \left( \bm{I} - \mathbf{n} \mathbf{n} \right) \cdot \mathbf{v}$ standing for the interfacial  velocity. Besides the standard diffusion and advection terms, Eq.~(\ref{advdiff}) also accounts for variations of surfactant concentration as a result of local changes in interfacial area.

Regarding the boundary conditions, all  fields  decay to their equilibrium values ($T_0$, $\mathbf{v}=\bm{0}$, $\Gamma_0$) far away from the disturbance. The boundary conditions on the bottom wall of the experimental cell actually depend on the specific situation and will be stipulated later.
At the liquid-air interface, conservation of momentum include  contributions from the viscous shear stress, Marangoni and capillary forces, as well as interfacial rheology~\cite{manikantanJFM2020}. The latter  can be neglected in the dilute phase of surfactant so that the stress balance condition reads
%%%%%%%%%%%%%%%%%%%%%%%%%%%%%%%%%%%%%%%%%%%
\begin{equation}
\bm{\sigma} \cdot \mathbf{n} = \bm{\nabla}_s \gamma +\gamma_0 \left( \bm{\nabla}_s \cdot \mathbf{n} \right) \mathbf{n} \ .
\label{stress}
\end{equation}
%%%%%%%%%%%%%%%%%%%%%%%%%%%%%%%%%%%%%%%%%%%
The cartesian components of the stress tensor $\bm{\sigma}$ are given by $\sigma_{ij} = -P \delta_{ij} + \eta ( \frac{\partial v_i}{\partial x_j} + \frac{\partial v_j}{\partial x_i})$. Note that, for the sake of convenience, the reference pressure of the gas phase is arbitrarily set to $0$. The first term on the right-hand side of Eq.~(\ref{stress}) is the Marangoni stress. To be consistent with Boussinesq approximation, the Laplace pressure term involves the equilibrium value~$\gamma_0$ of the surface tension. 
The normal fluid velocity is moreover related to the deformation rate of the interface through the kinematic condition
%%%%%%%%%%%%%%%%%%%%%%%%%%%%%%%%%%%%%%%%%%%
\begin{equation}
\mathbf{v} \cdot \mathbf{n} = -\frac{1}{\vert \bm{\nabla} \Phi \vert } \frac{\partial \Phi}{\partial t}   \ .
\label{kinematic}
\end{equation}
%%%%%%%%%%%%%%%%%%%%%%%%%%%%%%%%%%%%%%%%%%%
Finally, heat exchanges between the liquid and the gas phases are accounted for by Newton's law of cooling 
%%%%%%%%%%%%%%%%%%%%%%%%%%%%%%%%%%%%%%%%%%%
\begin{equation}
\kappa  \bm{\nabla} T \cdot \mathbf{n} + \beta \big( T_s-T_0 \big) =0  \ ,
\label{cooling}
\end{equation}
%%%%%%%%%%%%%%%%%%%%%%%%%%%%%%%%%%%%%%%%%%%
with $\beta$ the heat transfer coefficient and $T_s$ the temperature at the interface.

\section{Low-compressibility assumption}
\label{sec_inc}

Exact solutions to non-linear set of Eqs.~(\ref{momentum})--(\ref{cooling}) are rather scarce in the literature, with the notable exceptions of the pioneering work of Bratukhin and Maurin regarding surfactant-free thermocapillary flows~\cite{bratukhin1967}, or the recent advances on isothermal Marangoni spreading~\cite{crowdySIAM2021,bickelPRE2022,crowdyJFM2023,temprano2024}. Besides the non-linear nature of the mathematical problem, the complexity also arises from the large number of dimensionless groups that are involved in the physical description of the system. Still, there is a compelling need for theoretical predictions to assess the effect of surface-active contaminants on thermally-driven, as well as more general~\cite{bickelPRF2019,koleskiPRF2021}, interfacial flows. In order to make the problem analytically tractable, some simplifying assumptions are therefore required. Our goal in this section is to introduce and justify the low-compressibility approximation which is made  in order to linearize the mathematical problem. 

We first make the hypothesis that interface deformations remain sufficiently smooth so that the small-gradient approximation applies. More precisely, the interface profile is described by $\Phi(\mathbf{r},z,t)=z-h(\mathbf{r},t)$, where $h(\mathbf{r},t)$ denotes the height of the interface above the horizontal plane, $\mathbf{r}=(x,y)$ being the in-plane position vector. The unit normal vector thus reads
%%%%%%%%%%%%%%%%%%%%%%%%%%%%%%%%%%%%%%%%%%%
\begin{equation}
\mathbf{n}=\frac{1}{\sqrt{1 + \left( \bm{\nabla}_{\parallel} h\right)^2}} 
\begin{pmatrix}
-\frac{\partial h}{\partial x} \\
-\frac{\partial h}{\partial y} \\
1
\end{pmatrix}   \ ,
\end{equation}
%%%%%%%%%%%%%%%%%%%%%%%%%%%%%%%%%%%%%%%%%%%
where the 2D del operator is defined as $\bm{\nabla}_{\parallel}=(\frac{\partial}{\partial x},\frac{\partial }{\partial y})$. Here and in the following, the $\parallel$ subscript denotes projection onto the horizontal reference plane $xOy$.  We will subsequently assume that the gradients of surface deformations are small, namely $\vert \bm{\nabla}_{\parallel}h \vert \ll 1$. This assumption is all the more legitimate as interfacial deformations are excepted to be strongly suppressed in the presence of surfactants~\cite{vinnichenkoIJHMT2018,rudenkoJFM2022}.

We then wish to rewrite the transport equations in dimensionless form. 
To this aim, one can note that the heat source is featured by a typical temperature increase $\Delta T$ and a characteristic length scale $\ell$. 
The latter is to be identified with either the depth~$H$ [case~1], the radius~$a$ of the heat source [case~2], or the waist $r_0$ of the laser beam [case~3].
We can then construct two characteristic velocities related to the thermal actuation of the fluid. The thermocapillary velocity, $U_{\theta}=\gamma_{\theta} \Delta T/\eta$, is the velocity scale associated with the Marangoni flow due to the thermal gradient along the interface.  The second is the thermal buoyancy velocity, $U_{\alpha}=\rho_0g \alpha \ell^2 \Delta T/\eta$, which is related to the convection of the liquid as a result of the vertical thermal gradient. Equating both  then defines the thermal length scale
%%%%%%%%%%%%%%%%%%%%%%%%%%%%%%%%%%%%%%%%%%%
\begin{equation}
\ell_{\mathit th} = \left( \frac{\gamma_{\theta}}{\rho_0g\alpha }\right)^{1/2} \ . 
\end{equation}
%%%%%%%%%%%%%%%%%%%%%%%%%%%%%%%%%%%%%%%%%%%
For water at room temperature, one gets~$\ell_{\mathit th} \approx 1~\text{cm}$. 
The physical meaning  of $\ell_{\mathit th}$ is the following: when the characteristic length scale $\ell$ of the system is much larger than the thermal length, buoyancy is dominant and thermocapillary effects can be ignored. Otherwise, buoyancy  can be neglected.  In the following, emphasis is placed on situations where the thermocapillary effect is dominant: $\ell < \ell_{\mathit th}$.
The thermocapillary velocity~$U_{\theta}$ is therefore chosen to set the intensity of the primary flow that squeezes the surfactants.

From a thermodynamic viewpoint,  insoluble surfactants behave as a 2D gas with surface pressure $\Pi=\gamma_0-\gamma$. In order to assess the resistance of the surfactant-laden interface to compression, the surface compressibility~$\mathcal{B}$ is defined by analogy with bulk systems as~\cite{manikantanJFM2020}
%%%%%%%%%%%%%%%%%%%%%%%%%%%%%%%%%%%%%%%%%%%
\begin{equation}
\mathcal{B} = \frac{1}{\Gamma} \frac{\partial  \Gamma}{\partial  \Pi}\bigg\vert_{\Gamma=\Gamma_0} = \frac{1}{\gamma_s \Gamma_0} \ ,
\end{equation}
%%%%%%%%%%%%%%%%%%%%%%%%%%%%%%%%%%%%%%%%%%%
where the second equality follow from Eq.~(\ref{gamma}).  Notice that the surface compressibility is related to the velocity scale  $U_s = \gamma_s \Gamma_0/ \eta=(\eta \mathcal{B})^{-1}$ related with the solutocapillary effect.
For the ensuing analysis, it is appropriate to introduce the dimensionless compressibility according to
%%%%%%%%%%%%%%%%%%%%%%%%%%%%%%%%%%%%%%%%%%%
\begin{equation}
\varepsilon = \frac{\eta D_s }{\gamma_s \Gamma_0 \ell } = \text{Ma}_s^{-1} \ .
\end{equation}
%%%%%%%%%%%%%%%%%%%%%%%%%%%%%%%%%%%%%%%%%%%
So defined, the parameter $\varepsilon$ also corresponds to the inverse of the Marangoni number~$\text{Ma}_s$. We then follow the  line of reasoning of Shardt and collaborators~\cite{shardtJFM2016} and focus on the low-compressibility limit $\varepsilon \ll 1$. Indeed, compression of the surfactant layer drives a reverse Marangoni flow that resists the constraint. As a consequence, insoluble surfactants are almost always surface incompressible, even in the dilute regime~\cite{manikantanJFM2020}.  

It is thus assumed that deviations from the average concentration are very small: $\vert \Gamma - \Gamma_0 \vert \ll \Gamma_0$. 
Switching to non-dimensional variables, we are naturally led to set
%%%%%%%%%%%%%%%%%%%%%%%%%%%%%%%%%%%%%%%%%%%
\begin{equation}
c=\frac{\Gamma- \Gamma_0}{\varepsilon \Gamma_0} \ .
\end{equation}
%%%%%%%%%%%%%%%%%%%%%%%%%%%%%%%%%%%%%%%%%%%
The temperature is then made dimensionless according to   $\theta=(T-T_0)/\Delta T$.
Following~\cite{shardtJFM2016}, the reduction of the flow intensity is accounted for by introducing the dimensionless velocity $\mathbf{u}=(u,v,w)=\mathbf{v}/(\varepsilon U_{\theta})$. The dimensionless pressure is defined as  $p=\ell (P+\rho_0 gz)/ (\eta \varepsilon U_{\theta})$. 
From now on, coordinates are expressed in terms of the characteristic length scale $\ell$, and the time variable is given in units of the advection time scale  $\ell/ U_{\theta}$.
Finally, the position of the interface is expressed in units of $\varepsilon \ell \text{Ca}$, with $\text{Ca}=\eta U_{\theta} /\gamma_0$ the capillary number. With this choice, the small-gradient approximation is automatically satisfied since $\vert \bm{\nabla}_{\parallel} h \vert \sim \mathcal{O} ( \varepsilon )\ll 1$.

The transport problem defined by Eqs.~(\ref{momentum})--(\ref{cooling}) can thus be rewritten in dimensionless form. We moreover focus on the low-compressibility limit $\varepsilon \ll 1$ such that the terms of order $\mathcal{O}(\varepsilon)$ are systematically disregarded. This is in particular the case for the nonlinear term in the Navier-Stokes Eq.~(\ref{momentum}), that reads  for the dimensionless velocity field $\mathbf{u}(\mathbf{r},z,t)$ 
 %%%%%%%%%%%%%%%%%%%%%%%%%%%%%%%%%%%%%%%%%%%
\begin{equation}
\text{Re}\,  \frac{\partial \mathbf{u}}{\partial t}  =  \nabla^2 \mathbf{u}- \bm{\nabla} p + \mathcal{A} \theta \mathbf{e}_z  \ ,
\label{NS_adim}
\end{equation}
%%%%%%%%%%%%%%%%%%%%%%%%%%%%%%%%%%%%%%%%%%%
with $\text{Re}=\rho_0 U_{\theta} \ell/\eta$ the Reynolds number, and   $\mathcal{A}=U_{\alpha}/(\varepsilon U_{\theta})$ the dimensionless thermal expansion coefficient.
The continuity Eq.~(\ref{incomp})  remains unchanged
%%%%%%%%%%%%%%%%%%%%%%%%%%%%%%%%%%%%%%%%%%%
\begin{equation}
\bm{\nabla} \cdot \mathbf{u}=0  \ ,
\label{cont_adim}
\end{equation}
%%%%%%%%%%%%%%%%%%%%%%%%%%%%%%%%%%%%%%%%%%%
while the heat Eq.~(\ref{energy})  for the dimensionless temperature  $\theta(\mathbf{r},z,t)$ now becomes
%%%%%%%%%%%%%%%%%%%%%%%%%%%%%%%%%%%%%%%%%%%
\begin{equation}
\text{Pe}_{th}  \frac{\partial \theta}{\partial t}= \nabla^2 \theta + 
\begin{cases}
2 \pi \delta (x) \delta (z + 1 ) & \text{[case 1]} \ , \\
r^{-1} \delta (r) \delta (z) & \text{[case 2]} \ , \\
k e^{kz} e^{-r^2/2} & \text{[case 3]} \ .
\end{cases} 
\label{heat_adim}
\end{equation}
%%%%%%%%%%%%%%%%%%%%%%%%%%%%%%%%%%%%%%%%%%%
The thermal P\'eclet number is defined as $\text{Pe}_{th}=U_{\theta}\ell/D_{th}$, with $D_{th}=\kappa/(\rho_0 c)$ the thermal diffusivity.
Regarding the transport of surfactant, the terms related to surface curvature can be neglected in the small-gradient approximation. The advection-diffusion Eq.~(\ref{advdiff}) for the concentration field $c(\mathbf{r},t)$   eventually takes the dimensionless form
%%%%%%%%%%%%%%%%%%%%%%%%%%%%%%%%%%%%%%%%%%%
\begin{equation}
 \text{Pe}_{s} \left( \frac{\partial  c}{\partial t} + \bm{\nabla} _{\parallel} \cdot  \mathbf{u}_{\parallel}  \right)  =  \nabla_{\parallel}^2  c   \ ,
\label{advdiff_adim}
\end{equation}
%%%%%%%%%%%%%%%%%%%%%%%%%%%%%%%%%%%%%%%%%%%
where the solutal P\'eclet number $\text{Pe}_{s}=U_{\theta} \ell/D_{s}$ is defined accordingly.
We proceed likewise for the  interfacial boundary conditions~(\ref{stress})--(\ref{cooling}) at $z =0$ that simplify to 
%%%%%%%%%%%%%%%%%%%%%%%%%%%%%%%%%%%%%%%%%%%
\begin{subequations}
 \label{bc_adim}
\begin{align}
    &  p\big\vert_{z =0} - 2 \frac{\partial \mathbf{u}_{\parallel}}{\partial z}\bigg\vert_{z =0}  =  \bm{\nabla}^2_{\parallel}  h - \text{Bo}\, h  \ ,   \label{perpstress_adim} \\
    &\bm{\nabla} _{\parallel} c   =  -   \text{Pe}_s \bm{\nabla} _{\parallel} \theta\big\vert_{z =0}   \ , \label{parallelstress_adim} \\
    & w \big\vert_{z =0} = \frac{\partial h}{\partial t}  \ , \label{kinematic_adim} \\
    & \left( \frac{\partial  \theta }{\partial z}  + \text{Bi} \, \theta \right)\bigg\vert_{z =0} = 0  \ . \label{heatBC_adim}
\end{align}
\end{subequations}
%%%%%%%%%%%%%%%%%%%%%%%%%%%%%%%%%%%%%%%%%%%
Here, two additional dimensionless group are introduced: the Bond number $\text{Bo}= \rho_0 g \ell^2/\gamma_0$, which is the ratio of gravitational to capillary forces, and the Biot number $\text{Bi}=\beta\ell/\kappa$, which characterizes heat transfer at the interface.  The limit $\text{Bi} \to 0$ corresponds to the adiabatic (Neumann) boundary condition $\partial  \theta / \partial z \vert_{z=0}=0$, whereas the opposite limit $\text{Bi} \to \infty$ corresponds to the isothermal (Dirichlet) condition $\theta \vert_{z=0}=0$.
The condition~(\ref{parallelstress_adim}) is especially interesting: it reveals  that the solutal contribution  to the Marangoni stress exactly cancels the thermal contribution. The residual viscous stress being of higher order, it does not contribute to the stress balance at the interface.

The new mathematical problem defined by Eqs.~(\ref{NS_adim})--(\ref{bc_adim}) only involves \emph{linear} equations in low-compressibility limit. It should be emphasized that the only  small parameter that  was introduced so far is the inverse Marangoni number~$\varepsilon=\text{Ma}_s^{-1}$. No specific assumption has been made regarding the other dimensionless groups (Reynolds, thermal and solutal P\'eclet,  Bond, Biot, capillary), which can all be $\sim O(1)$. As a matter of fact, our description is appropriate as long as $\text{Ma}_s$ is much larger that the Reynolds and both P\'eclet numbers. These conditions are readily satisfied, even at very low surface concentration. The problem is therefore analytically tractable, at least in principle, in the low-compressibility limit~$\varepsilon \ll 1$.

\section{Flow properties in the steady-state}
\label{seq_steady}

Starting from an initial quiet situation, a Marangoni flow is triggered by switching on the heating. The system then reaches a steady-state which is the focus of the remaining of the paper.  The strategy is then the following: one first solves the  heat Eq.~(\ref{heat_adim}) in its stationary version $\nabla^2 \theta =- q(\mathbf{r},z)$, together with  Newton condition~(\ref{heatBC_adim}) at the interface. From this point, one can obtain  the surface concentration from parallel stress balance Eq.~(\ref{parallelstress_adim}). 
The problem is then closed thanks to the surfactant transport Eq.~(\ref{advdiff_adim}) that simplifies to $\bm{\nabla} _{\parallel} \cdot  \mathbf{u}_{\parallel}  = \text{Pe}_{s}^{-1} \nabla_{\parallel}^2  c$. This relation is important as it states that  a slight surface compressibility is needed for the mathematical problem to admit a non-trivial solution.

We can conclude from this short discussion that interfacial properties are actually determined in a straightforward manner. Note in passing that neither buoyancy nor capillarity are expected to matter regarding the interfacial flow. 
It requires some additional work, however, to determine the entire flow field in the bulk. This is achieved by solving Eq.~(\ref{NS_adim}) in steady-state, $\nabla^2 \mathbf{u}- \bm{\nabla} p + \mathcal{A} \theta \mathbf{e}_z =\bm{0}$,  together with the incompressibility condition Eq.~(\ref{cont_adim}) as well as the kinematic condition~(\ref{kinematic_adim}) that now reads $w \big\vert_{z =0}$. The deformation of the interface then follows  from the normal stress balance Eq.~(\ref{perpstress_adim}).

This general method  is now applied to three experimentally relevant situations.

\subsection{Heating by a line source}
\label{sec_line}

\begin{figure}
\centering
\includegraphics[width=\columnwidth]{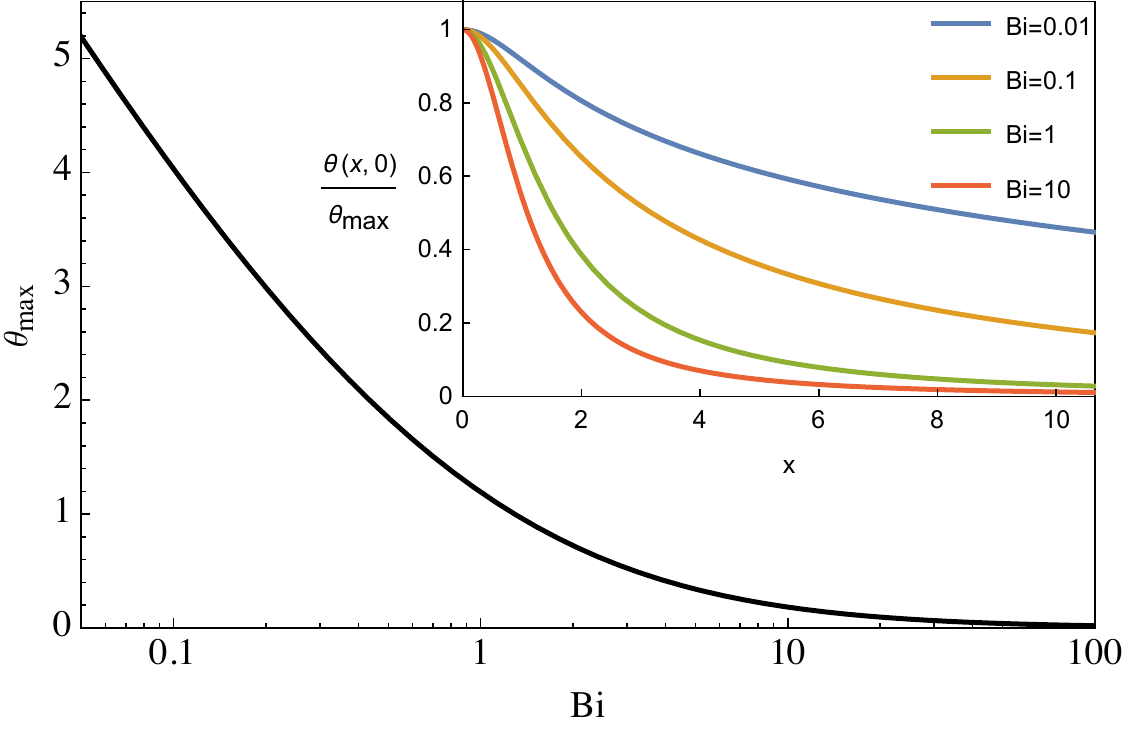}
\caption{Maximum  interfacial temperature $\theta_{\textrm{max}}=\theta(0,0)$ as  a function of the Biot number  (in logarithmic scale) for a line source. Inset: normalized interfacial temperature as a function of distance, for different values of $\text{Bi}$.}
\label{fig_temp2D}       
\end{figure}

As a first illustration, we consider the effect of airborne contaminants on the flow induced by a line heat source~[case~1]. In practice, this can be achieved by Joule effect using an electric wire immersed at some depth~$H$ below the free liquid surface~\cite{vinnichenkoIJHMT2018,rudenkoJFM2022,shahdhaarPoF2023}. The thickness of the wire is assumed to be much smaller than~$H$, which is naturally  identified as the reference length scale~$\ell$ of the model. We make no further assumption regarding buoyancy or capillary effects. Since the system is translationally invariant along the wire axis,  here defined  as the~$y$-direction, the problem is actually two-dimensional. 
It is then convenient to define the Fourier transform and its inverse as
%%%%%%%%%%%%%%%%%%%%%%%%%%%%%%%%%%%%%%%%%%%
\begin{subequations}
\begin{align}
& \tilde{f} (q) =\mathcal{F}[f(x)]=\int_{-\infty}^{+\infty}e^{-\mathrm{i}qx}f(x) \d x \ ,   \\
& f(x) =\mathcal{F}^{-1}[\tilde{f}(q)]= \frac{1}{2\pi} \int_{-\infty}^{+\infty}e^{\mathrm{i}qx}\tilde{f}(q) \d q \ .
\end{align}
\end{subequations}
%%%%%%%%%%%%%%%%%%%%%%%%%%%%%%%%%%%%%%%%%%%
The heat Eq.~(\ref{heat_adim}) then transforms into the following Green's function equation for the dimensionless temperature $\tilde{\theta}(q,z)=\mathcal{F}[\theta(x,z)]$
%%%%%%%%%%%%%%%%%%%%%%%%%%%%%%%%%%%%%%%%%%%
\begin{equation}
\frac{\partial^2  \tilde{\theta}}{\partial z^2}- q^2  \tilde{\theta} = - 2\pi \delta(z+1) \ .
\end{equation}
%%%%%%%%%%%%%%%%%%%%%%%%%%%%%%%%%%%%%%%%%%%
This equation, together with the cooling law~(\ref{heatBC_adim}), is readily solved. We moreover consider that the liquid subphase is semi-inifinite, and consequently assume that $\tilde{\theta}(q,z)$ vanishes when $z \to - \infty$. We then find
%%%%%%%%%%%%%%%%%%%%%%%%%%%%%%%%%%%%%%%%%%%
\begin{equation}
\tilde{\theta}(q,z) = \frac{\pi}{\vert q \vert} e^{-\vert q (z+1) \vert} + \frac{\pi}{\vert q \vert} \left(  \frac{\vert q \vert -\text{Bi}}{\vert q \vert + \text{Bi} }\right) e^{\vert q\vert (z-1)}  \ .
\label{tempfour11}
\end{equation}
%%%%%%%%%%%%%%%%%%%%%%%%%%%%%%%%%%%%%%%%%%%
This result can then be inverse Fourier transformed. However, since  our focus is on interfacial features, we give only the expression of the temperature field at $z = 0$
%%%%%%%%%%%%%%%%%%%%%%%%%%%%%%%%%%%%%%%%%%%
\begin{align}
\theta(x,0) =  e^{\text{Bi} (1+\mathrm{i}x) } & E_1\big[ \text{Bi}(1+\mathrm{i}x)\big] \nonumber \\
& + e^{\text{Bi} (1-\mathrm{i}x) } E_1\big[ \text{Bi}(1-\mathrm{i}x)\big] \ ,
\end{align}
%%%%%%%%%%%%%%%%%%%%%%%%%%%%%%%%%%%%%%%%%%%
with $E_1(z)= \int_z^{\infty} t^{-1} e^{-t} \d t $ the exponential integral function. The interfacial temperature is plotted  in Fig.~\ref{fig_temp2D}. 
As expected, its maximum   is reached just above the heat source: $\theta_{\textrm{max}}=\theta(0,0)=2e^{\text{Bi}} E_1(\text{Bi})$. In the limit $\text{Bi} \to \infty$, it is found to vanish as $\theta_{\textrm{max}} \sim 2/\text{Bi}$. Indeed,  the heat flux condition~(\ref{heatBC_adim}) simplifies to $\theta(x, 0)=0$  when $\text{Bi} \gg 1$. One therefore expects the interfacial temperature to become homogeneous in this limit. In the opposite regime $\text{Bi} \ll 1$, which corresponds to the insulating boundary condition $\partial \theta / \partial z \vert_{z = 0} = 0$,  the maximum temperature  diverges as $\theta_{\textrm{max}} \sim - 2 \ln (\text{Bi})$. The Biot number should therefore be kept finite in order to avoid mathematical singularities. The width of the temperature profile is also shown in inset of Fig.~\ref{fig_temp2D}. One observes that the temperature decays very slowly in the adiabatic limit ($\text{Bi} \ll 1$),  whereas the distribution becomes sharper as    $\text{Bi}$ increases. Far away from the heat source ($x \gg 1)$, the interfacial temperature is found to decay like $\theta(x,0) \sim x^{-2}$.

%%%%%%%%%%%%%%%%%%%%%%%%%%%%%%%%%%%%%%%
%%%%%%%%%%%%%%%%%%%%%%%%%%%%%%%%%%%%%%%

The discussion regarding the concentration  closely follows that of the temperature since the integration of Eq.~(\ref{parallelstress_adim}) directly gives 
%%%%%%%%%%%%%%%%%%%%%%%%%%%%%%%%%%%%%%%%%%%
\begin{equation}
c(x) = -\text{Pe}_s \theta(x,0) \ .
\end{equation}
%%%%%%%%%%%%%%%%%%%%%%%%%%%%%%%%%%%%%%%%%%%
We find in particular that surfactants are depleted from the vicinity of the origin.  Finally, the interfacial velocity is obtained by integration of Eq.~(\ref{advdiff_adim}), that now  reads
%%%%%%%%%%%%%%%%%%%%%%%%%%%%%%%%%%%%%%%%%%%
\begin{equation}
\frac{\partial u}{\partial x} \bigg\vert_{z=0} = \text{Pe}_{s}^{-1} \frac{\partial^2 c}{\partial x^2} \ .
\end{equation}
%%%%%%%%%%%%%%%%%%%%%%%%%%%%%%%%%%%%%%%%%%%
It is then straightforward to obtain the interfacial velocity in Fourier representation, $\tilde{u}(q,0)=-iq \tilde{\theta}(q,0)$, with $\tilde{\theta}(q,0)$  given in Eq.~(\ref{tempfour11}).
At this point one still needs to perform the inverse Fourier transform. This is readily achieved in the adiabatic limit $\text{Bi} \to 0$: even though the temperature diverges logarithmically in the vicinity of the origin, the behavior of the velocity field happens to be regular. One finally gets for $\text{Bi}= 0$
%%%%%%%%%%%%%%%%%%%%%%%%%%%%%%%%%%%%%%%%%%%
\begin{equation}
u(x,0) = \frac{2 x}{1+x^2}  \ .
\end{equation}
%%%%%%%%%%%%%%%%%%%%%%%%%%%%%%%%%%%%%%%%%%%
The interfacial velocity therefore vanishes like $u(x,0)\sim x^{-1}$ at large distances $x \gg 1$.

\subsection{Heating by a point-like source: 3D axisymmetric flow}
\label{sec_point}

We now focus on the 3D axisymmetric flow that results from the local heating of the water-air interface~[case~2]. This configuration was the subject of several experimental works in recent years~\cite{girotLangmuir2016,koleskiPoF2020,pinanPoF2021}, with the general conclusion that the flow features strongly differ from what is expected for a clean interface. In the theoretical description, the reference length scale $\ell$ has to be identified with the actual size $a$ of the source, which experimentally can be as small as a few microns. For the sake of simplicity, we directly assume an adiabatic boundary condition ($\text{Bi}=0$) at the liquid-air interface. We also make the hypothesis that the liquid phase is semi-inifinite.

Given the cylindrical symmetry of the problem, it is convenient to define the Hankel transforms  according to~\cite{piessensHankelbook2000}
%%%%%%%%%%%%%%%%%%%%%%%%%%%%%%%%%%%%%%%%%%%
\begin{equation}
\breve{f}(q) =\mathcal{H}_{\nu}\left[ f(r)\right] =\int_{0}^{+\infty} f(r) J_{\nu}(qr) r \d r \ , \end{equation}
%%%%%%%%%%%%%%%%%%%%%%%%%%%%%%%%%%%%%%%%%%%
and its inverse
%%%%%%%%%%%%%%%%%%%%%%%%%%%%%%%%%%%%%%%%%%%
\begin{equation}
f(r) =\mathcal{H}^{-1}_{\nu}\left[ f_{\nu}(q)\right]=  \int_{0}^{+\infty} \breve{f}(q)  J_{\nu}(qr) q\d q \ ,
\end{equation}
%%%%%%%%%%%%%%%%%%%%%%%%%%%%%%%%%%%%%%%%%%%
where $J_{\nu}$ is the Bessel function of order $\nu$. The temperature $\breve{\theta}=\mathcal{H}_{0}[\theta]$  then satisfies the Hankel transform of the heat Eq.~(\ref{heat_adim})
%%%%%%%%%%%%%%%%%%%%%%%%%%%%%%%%%%%%%%%%%%%
\begin{equation}
\frac{\partial^2 \breve{\theta}}{\partial z^2}- q^2 \breve{\theta} = -  \delta(z) \ .
\end{equation}
%%%%%%%%%%%%%%%%%%%%%%%%%%%%%%%%%%%%%%%%%%%
Enforcing the boundary conditions $\partial \breve{\theta} / \partial z \vert_{z=0} = 0$ and $\breve{\theta}(q,z) \to 0$ when $z \to -\infty$, one obtains  
%%%%%%%%%%%%%%%%%%%%%%%%%%%%%%%%%%%%%%%%%%%
\begin{equation}
\breve{\theta}(q,z) = \frac{e^{qz}}{q} \ .
\label{tempstat}
\end{equation}
%%%%%%%%%%%%%%%%%%%%%%%%%%%%%%%%%%%%%%%%%%%
The expression of the Hankel transform $\breve{c}=\mathcal{H}_{0}[c]$ of the concentration then follows from Eq.~(\ref{parallelstress_adim}) and we find 
%%%%%%%%%%%%%%%%%%%%%%%%%%%%%%%%%%%%%%%%%%%
\begin{equation}
\breve{c}(q) = - \frac{\text{Pe}_s}{q} \ .
\label{cstat}
\end{equation}
%%%%%%%%%%%%%%%%%%%%%%%%%%%%%%%%%%%%%%%%%%%

%%%%%%%%%%%%%%%%%%%%%%%%%%%%%%%%%%%%%%%
\begin{figure}
\centering
\includegraphics[width=\columnwidth]{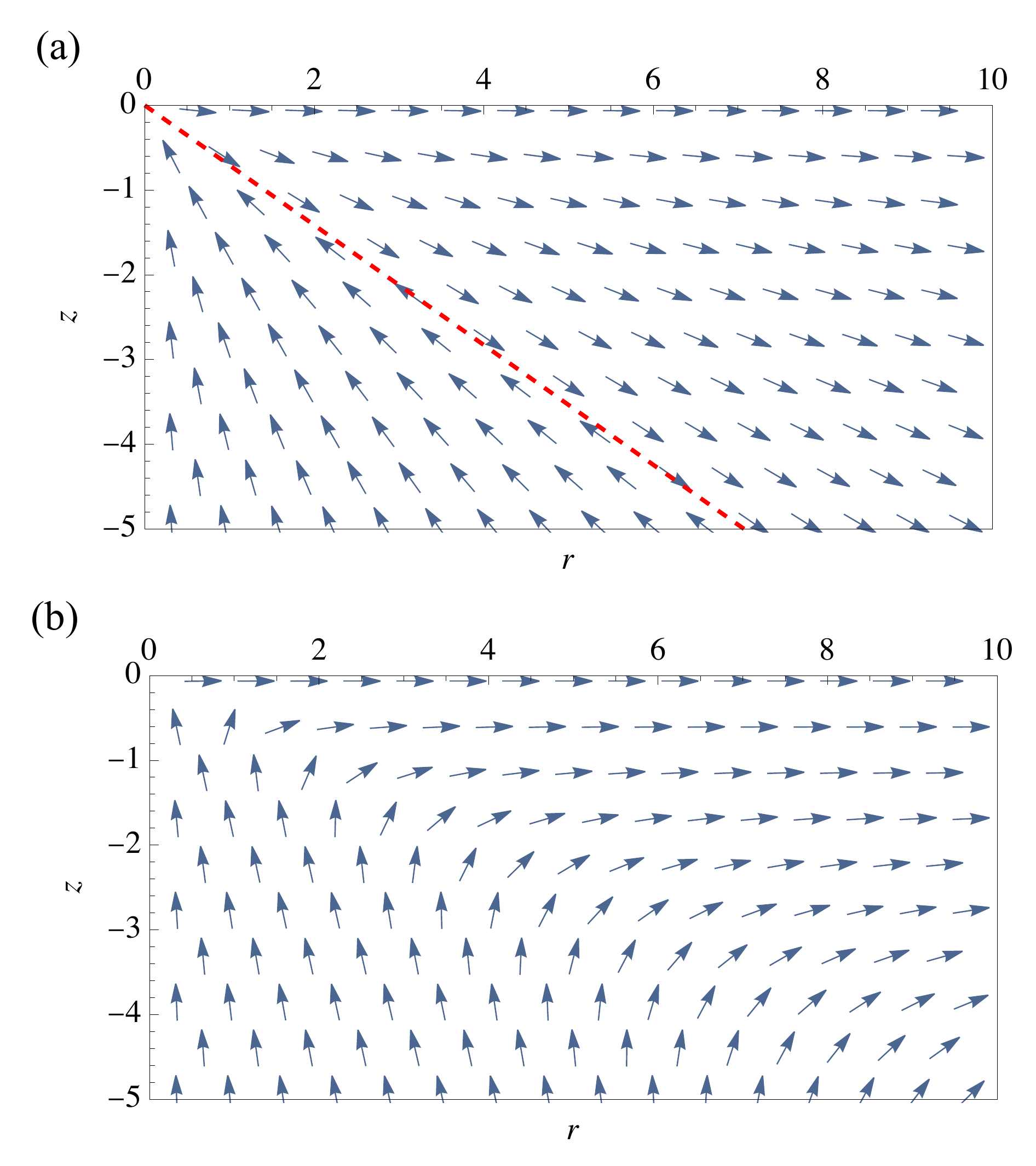}
\caption{Streamlines of the 3D  flow due to a point-like heat source for (a) surfactant-laden \textit{vs.} (b) clean interface (for $\text{Re}=0$).}
\label{fig_flow2}       
\end{figure}
%%%%%%%%%%%%%%%%%%%%%%%%%%%%%%%%%%%%%%%

The interfacial velocity can be determined accordingly, yet it is interesting to characterize the flow field in the bulk as well. To this aim, it is assumed that  the velocity vanishes when $z \to -\infty$. The effect of buoyancy is neglected.
The incompressible flow Eqs.~(\ref{NS_adim}) and~(\ref{cont_adim})  can be rewritten for the radial component $\breve{u}=\mathcal{H}_{1}[ u]$ and the vertical component $\breve{w}=\mathcal{H}_{0}[ u]$ of the velocity field as
%%%%%%%%%%%%%%%%%%%%%%%%%%%%%%%%%%%%%%%%%%%
\begin{equation}
\frac{\partial^4 \breve{w}}{\partial z^4} -2q^2 \frac{\partial^2 \breve{w}}{\partial z^2} +q^4 \breve{w}=0 \ , 
\end{equation}
%%%%%%%%%%%%%%%%%%%%%%%%%%%%%%%%%%%%%%%%%%%
and
%%%%%%%%%%%%%%%%%%%%%%%%%%%%%%%%%%%%%%%%%%%
\begin{equation}
q \breve{u} + \frac{\partial \breve{w}}{\partial z} = 0 \ ,
\end{equation}
%%%%%%%%%%%%%%%%%%%%%%%%%%%%%%%%%%%%%%%%%%%
Solving these equations and enforcing the boundary condition at the interface $\breve{u}\vert_{z=0}=-q\text{Pe}_s^{-1} \breve{c}(q)=1$ and $\breve{w}\vert_{z=0}=0$ finally leads to
%%%%%%%%%%%%%%%%%%%%%%%%%%%%%%%%%%%%%%%%%%%
\begin{equation}
\breve{u}(q,z) = \left( 1 +q z \right) e^{qz} \ , \quad \text{and} \quad \breve{w}(q,z) = -q z  e^{qz} \ .
\label{vstat}
\end{equation}
%%%%%%%%%%%%%%%%%%%%%%%%%%%%%%%%%%%%%%%%%%%
Computing the inverse Hankel transforms is then a straightforward operation. We thus predicts that both the temperature and  the concentration decay like the inverse of the distance  
%%%%%%%%%%%%%%%%%%%%%%%%%%%%%%%%%%%%%%%%%%%
\begin{equation}
T(r,z) = \frac{1}{\sqrt{r^2 + z^2}} \ , \quad \text{and} \quad c(r)  = - \frac{\text{Pe}_s}{r} \ .
\end{equation}
%%%%%%%%%%%%%%%%%%%%%%%%%%%%%%%%%%%%%%%%%%%
Regarding the velocity field, one obtains 
%%%%%%%%%%%%%%%%%%%%%%%%%%%%%%%%%%%%%%%%%%%
\begin{subequations}
\label{vel_case2}
\begin{align}
& u(r,z) =  \frac{r \left(r^2-2z^2\right)}{\left(r^2 + z^2\right)^{5/2}}    \ , \label{ux} \\
& w(r,z) =   \frac{z \left(r^2-2z^2\right)}{\left(r^2 + z^2\right)^{5/2}}   \ . \label{uz}
\end{align}
\end{subequations}
%%%%%%%%%%%%%%%%%%%%%%%%%%%%%%%%%%%%%%%%%%%
The flow streamlines are plotted in Fig.~\ref{fig_flow2}. For comparison, we also show the thermocapillary flow field for a ``clean'' interface (\textit{i.e.}, in the absence of surfactant) at vanishing Reynolds number~\cite{bratukhin1967,koleskiFluids2020}. Note that, in a real experiment, the streamlines are expected  to recirculate far way from the origin. This is due either to transient effects or to the finite size of the system, which are both ignored in this work. The present Fig.~\ref{fig_flow2} should therefore be understood as a ``zoom''  close to the origin.
Besides the different power-law, one clearly notices that the topology of the hindered flow is quite different. Indeed, if the pure thermocapillary flow has the topology of a vortex ring, this is not the case anymore when surfactants are present. In the latter case, the velocity actually cancels on a cone of equation $r=\sqrt{2} \vert z\vert$.  Experimental observation of this feature would be an undeniable marker of the presence of surfactants.

\subsection{Heating by a collimated laser beam}

%%%%%%%%%%%%%%%%%%%%%%%%%%%%%%%%%%%%%%%
\begin{figure}
\centering
\includegraphics[width=\columnwidth]{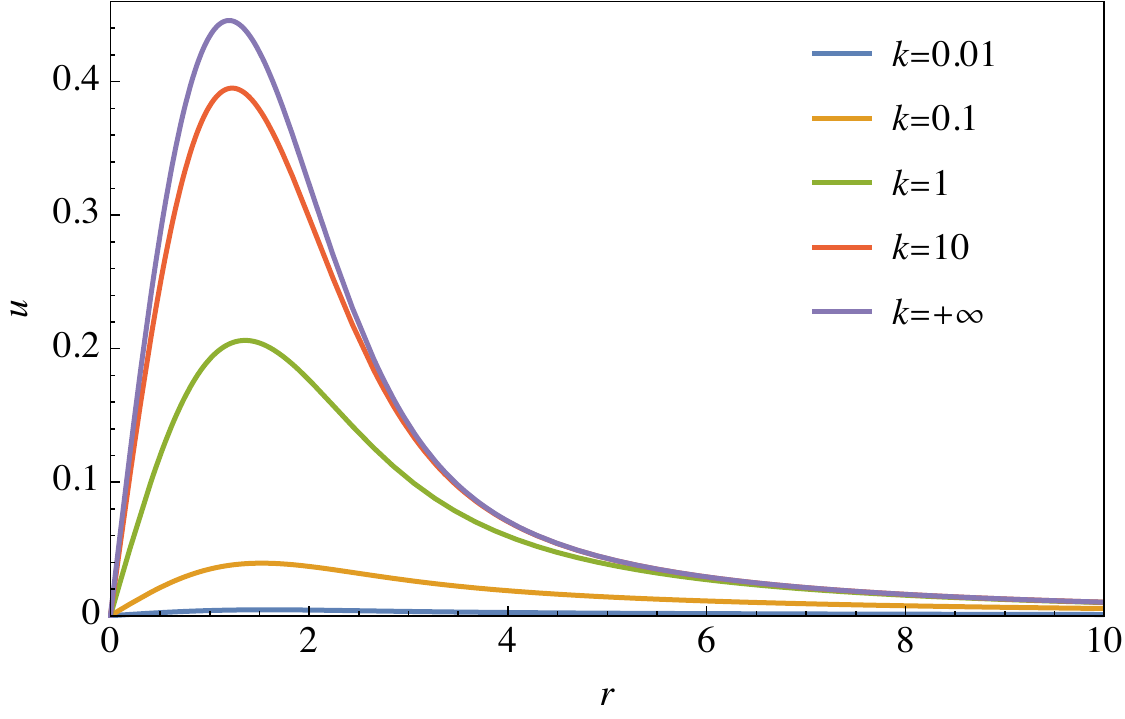}
\caption{Radial velocity field as a function of distance, for different values of the dimensionless laser penetration length.}
\label{fig_flow3}       
\end{figure}
%%%%%%%%%%%%%%%%%%%%%%%%%%%%%%%%%%%%%%%

As a third example, we consider the thermocapillary flow generated by a Gaussian laser beam~[case~3], under the assumption of a constant beam waist. We also account for light attenuation in the liquid phase~\cite{pinanPoF2021}. The temperature $\breve{\theta}=\mathcal{H}_{0}[\theta]$   satisfies the Hankel transform of the heat Eq.~(\ref{heat_adim})
%%%%%%%%%%%%%%%%%%%%%%%%%%%%%%%%%%%%%%%%%%%
\begin{equation}
\frac{\partial^2 \breve{\theta}}{\partial z^2}- q^2 \breve{\theta} = - ke^{kz} e^{-q^2/2} \ .
\end{equation}
%%%%%%%%%%%%%%%%%%%%%%%%%%%%%%%%%%%%%%%%%%%
For a semi-infinite liquid layer, the temperature is expected to vanish at infinity: $\breve{\theta}(q,z) \to 0$ when $z \to -\infty$.
Enforcing moreover the insulating boundary condition $\partial \breve{\theta} / \partial z \vert_{z=0} = 0$ at the interface, one finds  
%%%%%%%%%%%%%%%%%%%%%%%%%%%%%%%%%%%%%%%%%%%
\begin{equation}
\breve{\theta}(q,z) = \frac{k^2}{k^2-q^2} \left(\frac{e^{qz}}{q} -\frac{e^{kz}}{k} \right)  e^{-q^2/2} \ .
\label{tempstat3}
\end{equation}
%%%%%%%%%%%%%%%%%%%%%%%%%%%%%%%%%%%%%%%%%%%
The  radial velocity $\breve{u}=\mathcal{H}_{1}[ u]$ at the interface follows in a straightforward manner: $\breve{u}(q,0) = k e^{-q^2/2}/(k+q)$.
The inverse Hankel transform can then be  evaluated in the limit of large penetration length ($z_0 \gg r_0$ or $k \ll 1$)
%%%%%%%%%%%%%%%%%%%%%%%%%%%%%%%%%%%%%%%%%%%
\begin{equation}
u(r,0) = k\, \frac{1-e^{-r^2/2}}{r} \ ,
\end{equation}
%%%%%%%%%%%%%%%%%%%%%%%%%%%%%%%%%%%%%%%%%%%
as well as in the opposite limit of small penetration length ($z_0 \ll r_0$ or $k\gg 1$)
%%%%%%%%%%%%%%%%%%%%%%%%%%%%%%%%%%%%%%%%%%%
\begin{equation}
u(r,0) = \sqrt{\frac{\pi}{8}} re^{-r^2/4} \left[ I_0 \left( \frac{r^2}{4} \right) - I_1 \left( \frac{r^2}{4} \right)  \right] \ ,
\end{equation}
%%%%%%%%%%%%%%%%%%%%%%%%%%%%%%%%%%%%%%%%%%%
For intermediate values of $k$, one has to resort to numerical integration.  The result is plotted in Fig.~\ref{fig_flow3}. The velocity always vanishes at the origin, as expected from the radial symmetry. It first increases linearly at short distances until reaching a maximum at $r_{\textrm{max}} \approx 2r_0$. The amplitude of this maximum strongly depends on the penetration length. The velocity then decays to zero at larger distances.  One can note that our predictions are very similar to what was recently observed both experimentally and numerically (see, \textit{e.g.}, Fig.~3 of Ref.~\cite{pinanPoF2021}).

\section{Discussion}
\label{sec_disc}

To summarize, we have developed a general framework in order to account for the presence of a low concentration of surfactants. The thermocapillary flow is expected to be strongly hindered due to the counterflow that originates from the compression of the surfactant layer. As commonly observed in experiments, even a very low concentration of surface-active impurities can completely modify the hydrodynamic response of the interface. The purpose of this work was to provide a quantitative framework in order to assess the physical properties of these contaminants. 

The balance between thermocapillary and solutocapillary effect is expressed by the Marangoni boundary condition Eq.~(\ref{parallelstress_adim}), which reveals that the concentration adapts to the temperature gradient in order to cancel out the total stress. As a consequence, the shear contribution is irrelevant in the low-compressibility limit. The  velocity field is then set in a second step by the balance between advection and diffusion of surfactants.
We have also  shown that, in steady-state, neither buoyancy nor capillarity are expected to be relevant as far as interfacial quantities are concerned. Indeed, both the surface concentration and the interfacial velocity or solely driven by the interfacial temperature. This conclusion is of significance since most experiments actually probe interfacial flows. Interface deformations or the effect thermal buoyancy can nevertheless be readily elucidated, if relevant, as the transport equations are linear.

We now illustrate on a specific instance how quantitative information can be extracted from experimental data. For the discussion that follows, we  restore the dimensions of the physical quantities.
The focus is on the thermocapillary flow due to a point-like heat source that lies at the water-air interface [case~2]. For a ``clean'' interface, Bratukhin and Maurin~\cite{bratukhin1967} showed that the interfacial velocity~$v_r(r,0)$  decays with the distance~$r$ to the heat source as $v_r(r,0) \sim r^{-1}$. Interestingly, this scaling is valid for any Reynolds and thermal P\'eclet numbers~\cite{bratukhin1967}. Yet this simple behavior is hardly observed in experiments. Let us indeed consider a tracer particle at the liquid-air interface which is advected by the flow. If $R(t)$ denote the tracer position, its velocity is related to the interfacial velocity field through $\d R / \d t = v_r(R(t),0)\sim 1/R(t)$. 
Integrating this relation, one  then expects $R^2(t)$ to vary linearly with time, which is definitely in contradiction with the experimental data shown in  Fig.~7 of Ref.~\cite{koleskiPoF2020}. Instead, one clearly observes   for several tracers a linear variation of the cube of the distance: $R_{\textrm{exp}}^3(t)=At + B$, with $A \approx 2.7\times 10^{-9}~\text{m}^3 \, \text{s}^{-1}$. To understand this discrepancy, one has to account for the presence of surface-active impurities with an unknown concentration $\Gamma_0$. The interfacial velocity then scales as $v_r(r,0) \sim r^{-2}$. Restoring the prefactors, it can be deduced from Eq.~(\ref{vel_case2}) that
%%%%%%%%%%%%%%%%%%%%%%%%%%%%%%%%%%%%%%%%%%%
\begin{equation}
v_r(r,0)  = \frac{\gamma_{\theta} \Delta T}{\gamma_s \Gamma_0} \,  \frac{a D_s}{r^2} \ .
\label{v_dim}
\end{equation}
%%%%%%%%%%%%%%%%%%%%%%%%%%%%%%%%%%%%%%%%%%%
The hindered velocity is thus proportional to the temperature gradient $\Delta T$, that sets the fluid in motion. It is also  inversely proportional to the surface concentration~$\Gamma_0$, as the presence of surfactant counteracts the primary flow. Interestingly, the viscosity does not enter the expression of the interfacial velocity: this illustrates tha the residual flow is not driven by interfacial shear stress.
The equation of motion $\d R / \d t = v_r(R(t),0)$ can then be integrated  with the initial condition $R(t_0)=R_0$, and we find
%%%%%%%%%%%%%%%%%%%%%%%%%%%%%%%%%%%%%%%%%%%
\begin{equation}
R(t)^3 = R_0^3 + 3 \frac{\gamma_{\theta} \Delta T}{\gamma_s \Gamma_0}  aD_s (t-t_0) \ .
\end{equation}
%%%%%%%%%%%%%%%%%%%%%%%%%%%%%%%%%%%%%%%%%%%
This behavior now correctly matches  the experimental observations~\cite{koleskiPoF2020}. We can  get some additional insight regarding the concentration of surfactants. For $a\approx 100~\mu\text{m}$, $\Delta T \approx10~\text{K}$, $\gamma_{\theta} \approx 10^{-4}~\text{J}\, \text{m}^{-2} \, \text{K}^{-1}$, $\gamma_s \approx k_B T_0 \approx 10^{-21}~\text{J}$, and assuming a typical value $D_s \approx 10^{-9}~\text{m} \, \text{s}^{-2}$, one gets $\Gamma_0 \approx 100$~molecules~$\mu \text{m}^{-2}$. Even at such small concentration, the low-incompressibility hypothesis is holding since $\varepsilon \approx 10^{-1}$ for $\eta \approx 10^{-3}~\text{Pa} \, \text{s}$. Our analysis thus provides a quantitative estimate of the amount of surface-active impurities.

To close this discussion, we note that the predictions obtained from standard stagnant cap approach~\cite{bickelEPJE2019} are not compatible with the experimental observations. Indeed, if the diffusive term is neglected in Eq.~(\ref{advdiff}) from the beginning (limit $\text{Pe}_s \to \infty$), the velocity field is expected to decay like $v_r(r,0) \sim r^{-1}$ near the heat source and then to vanish at a finite distance. Instead, the scaling $v_r(r,0) \sim r^{-2}$ predicted in this work and observed in the experiments arises from the balance between advection and diffusion in the transport Eq.~(\ref{advdiff_adim}). It therefore requires the P\'eclet number to be finite, as further evidenced by the relation $v_r \propto \text{Pe}^{-1}_s$ obtained in Eq.~(\ref{v_dim}).

We conclude by stating that this work should open new perspectives regarding the effect of surface-active impurities on interfacial flows. The approach that we have developed following Ref.~\cite{shardtJFM2016} can readily be adapted to various experimental situations. The fact that the transport equations become linear in the low-compressibility limit  
allows for a systematic characterization of the physical properties of the contaminants. In particular, the relaxation dynamics after switching off the heating is expected to provide additional valuable information regarding the physical properties of the interfacial contaminants~\cite{koleskiPoF2020}.

%% End of file `jfm2esam.bib'.

\end{document}